\documentclass[12pt]{article}

\begin{document}
\thispagestyle{empty}
\title{\bf The problem of exotic states:\\ view from complex angular
momenta }
\author{Ya.~I. AZIMOV$^1$\\
\\
$^1${\it Petersburg Nuclear Physics Institute,} \\
{\it Gatchina, Leningrad area,} \\
{\it 188300, Russia}}
\date{}
\maketitle

\begin{abstract}
Having in mind present uncertainty of the experimental
situation in respect to exotic hadrons, it is important to discuss
any possible theoretical arguments, pro and contra. Up to now,
there are no theoretical ideas which could forbid existence of the
exotic states. Theoretical proofs for their existence are also
absent. However, there are some indirect arguments for the latter
case. It will be shown here, by using the complex angular momenta
approach, that the standard assumptions of analyticity and
unitarity for hadronic  amplitudes lead to a non-trivial
conclusion: the S-matrix has infinitely many poles in the energy
plane (accounting for all its Riemann sheets). This is true for
any arbitrary quantum numbers of the poles, exotic or non-exotic.
Whether some of the poles may provide physical (stable or
resonance) states, should be determined by some more detailed
dynamics.
\end{abstract}

\section {Exotic hadrons: brief overview}

The problem of exotic states of hadrons has long history, nearly
as long as quark themselves. After several unsuccessful
experimental searches, it was formulated most clearly by
Lipkin\cite{lip}, as a problem for theorists: ``Why are there no
strongly bound exotic states ..., like those of two quarks and two
antiquarks or four quarks and one antiquark?'' This question could
be eliminated by experimental observations of the first exotic
baryon $\Theta^+$, just consisting of at least four quarks (two
$u$'s and two $d$'s) and one antiquark ($\bar s$). However, other
experimental publications, with null results, cast doubts on its
existence (recent history and present experimental status see in
the talk~\cite{bur}). Therefore, Lipkin's question may be still
burning, having no answer. It seems reasonable in such a situation
to analyze in detail all arguments, pro and contra exotic hadrons.
Here we briefly discuss some of them.

The current experimental situation is rather uncertain and not
decisive yet. In the plenary talk of Burkert at NSTAR2005, it
was summarized by the words: ``The narrow $\Theta^+$ pentaquark
is not in good health, but it is too early to pronounce it dead''.

What about theoretical situation, the general postulates of
Quantum Field Theory (QFT) do not provide any way to discriminate
between exotic and non-exotic particles. Its particular case,
Quantum Chromodynamics (QCD), believed to underlie the strong
interactions and hadron spectroscopy, also cannot forbid exotic
hadrons {\it vs.} non-exotic ones (at least, at the present level
of understanding the structure and properties of QCD).

Moreover, in the framework of QCD, any hadron should be seen in
some conditions ({\it e.g.}, for short time intervals) as a
multi-quark system. Experiments on hard processes confirm this.
But it is then difficult to understand why such multi-quark
systems must be bound in all cases to have quantum numbers of
a 3-quark (or quark-antiquark) system.

Attempts to calculate hadronic spectra in various approaches,
which are assumed to be based on QCD (bag model, soliton-like
models, sum rules, lattice calculations, and so on), as a rule,
also demonstrate exotic states (though with their properties
strongly dependent on the model used).

Rather unexpectedly, the method of complex angular momenta (CAM),
usually related only to high-energy asymptotics of hadronic
amplitudes, has also something to say in the problem of exotic
states. It can suggest a new (indirect) argument for existence
of exotic hadrons. This argument has been recently
published~\cite{aaswg}. Here it will be explained in more detail.

\section{CAM and exotics}

Let us begin with some necessary preliminaries, partly forgotten
now. For simplicity, at first we neglect particle spins.

Consider a process 2-hadrons-into-2-hadrons. Its amplitude $A$ is
a function of two independent variables, for which one may choose
the c.m. energy $W$ and c.m. scattering angle $\theta$. Another
possibility is to use invariant variables. There are three of them
(Mandelstam variables): one is the c.m. energy squared $s=W^2$;
two others are squares of two c.m. momentum transfers $t$ and $u$,
between a particular initial hadron and one or the other final
hadron. Usually, $t$ is taken as proportional to $z=\cos\theta$,
with $u$ proportional to $-z$. Evidently, they are not
independent. Moreover, the sum $s+t+u$ equals just to the sum of
the squared masses for two initial and two final particles (being
thus independent of both $W$ and $z$).

The amplitude $A(s,z)$, as function of $z$,  may be decomposed into
partial waves. Every partial-wave amplitude $f_l(s)$ corresponds to
a definite value $l$ of the orbital momentum. For the physical
amplitudes, $l$ may only be equal to a non-negative integer number.
In the case of purely elastic scattering, the physical amplitudes
$f_l(s)$ satisfy the elastic unitarity condition
\begin{equation} \label{eq:p-u}
f_l(s)-f^*_l(s)=2ik\,f_l(s)\,f^*_l(s)\,,
\end{equation}
where $k$ is the c.m. momentum.

Now we make two assumptions:

1) Amplitudes $f_l(s)$ admit unambiguous analytical continuation
to non-integer, and even complex, values of $l$.

2) There are no massless hadrons and no massless hadron exchanges.

The first assumption is not trivial. Every function, defined on a
set of discreet points, can be analytically continued, but the
continuation is, generally, very ambiguous. Only in some cases
there exists a preferred continuation, which may be clearly
separated from all others. This is fulfilled, {\it e.g.}, if the
amplitude $A(s, z)$ satisfies dispersion relations (DR) in
momentum transfers $t,u$; the arising continuation is described by
the integral Gribov-Froissart (GF) formula\cite{gf} (for more
details see the monograph~\cite{col}). Such DR have never been
formally proved, neither in general QFT, nor in QCD. Nevertheless,
they (and their analogs) are widely and actively used in
phenomenology of strong interactions. Up to now, they have not
encountered any inconsistency. Note that the DR provide a
sufficient condition for the unambiguous continuation; necessary
conditions are essentially weaker.

The second assumption ensures a finite range of interactions, and
also the threshold behavior $\,\sim k^{2l}$ for the elastic
scattering amplitudes $f_l(s)$ with physical values of $l$, when
$k\to0$. The GF formula, where it is applicable, provides the same
threshold behavior for the continued amplitudes $f_l(s)$. But the
elastic unitarity condition for the continued amplitudes has
somewhat modified form
\begin{equation} \label{eq:c-u}
f_l(s)-f^*_{l^*}(s)=2ik\,f_l(s)\,f^*_{l^*}(s)\,,
\end{equation}
which coincides with eq.(\ref{eq:p-u}) only at real $l$.

It is easy to see now that the continued unitarity relation
(\ref{eq:c-u}) is not always consistent with the threshold
behavior $f_l(s)\sim k^{2l}$. Indeed, near threshold, each of the
left-hand side terms is $\,\sim k^{2{\rm Re}\,l}$, while the
right-hand side is $\,\sim k^{4{\rm Re}\,l+1}$. Since the
left-hand side terms may subtract each other, but not enhance, the
two sides may be consistent with each other only at ${\rm
Re}\,l>-1/2$. This problem was first discovered by Gribov and
Pomeranchuk\cite{gp}. To solve it, they studied in more detail the
small-$k^2$ region and showed that there are Regge poles, which
condense near threshold to the point $l=-1/2$ and, thus,
invalidate the $k^{2l}$-behavior of $f_l(s)$ for ${\rm
Re}\,l<-1/2$. Near the threshold, these poles have trajectories
\begin{equation}
\label{eq:grp} l_n(s)\approx -\frac12 + \frac{i\pi
n}{\ln(R\sqrt{|k^2|})} \,, \end{equation} with $R$ being the
effective interaction radius. The number $n$ takes any positive
and/or negative integer values, $n=\pm1,\pm2,...\,$. Hence, there
are infinitely many reggeons condensing to $l=-1/2$ at $k^2\to0$.

Till now, we have neglected particle spins. Taking them into
account changes the orbital momentum $l$ by the total angular
momentum $j$. The reggeon condensation still exists at a
two-particle threshold, though with the shifted limiting
point~\cite{az}. For the spins $\sigma_1$ and $\sigma_2$ it is
\begin{equation}\label{eq:jsp}
j=-1/2+\sigma_1+ \sigma_2\,, \end{equation} instead of $j=-1/2$,
without spins. The reason is simple: the condensation point, as
before, corresponds to $l=-1/2$, and the highest value of $j$ at
fixed $l$ equals $l+\sigma_1+\sigma_2$. The movement of the
condensing reggeons in the $j$-plane, near a threshold energy for
two spinning particles, is also described by trajectories
(\ref{eq:grp}), but with the shifted limiting point
(\ref{eq:jsp}).

The main conclusion of the above consideration is that the unitarity
and the possibility of unambiguous analytical continuation in $j$
(standard analytical properties, in particular), taken together, imply
existence of the Gribov-Pomeranchuk (GP) threshold condensations of
reggeons. They collect infinite number of reggeons and, therefore,
imply that the total number of reggeons is always infinite.

The reggeon positions depend on energy and are determined by a
relation of the form $F(s,\,j)=0\,$. Each of its solution
corresponds to an amplitude pole, which may be considered either
as a pole in $j$, with position depending on the energy (on $s$),
or as a pole in the energy (in $s$), with position depending on
the total angular momentum $j$. This provides one-to-one
correspondence between reggeons and spin-dependent poles in the
energy plane. Therefore, the infinite number of reggeons
corresponds to the infinite number of poles in the energy plane.

Note that no assumptions about quantum numbers of the poles have
been used. Hence, the hadronic $S$-matrix should have infinite
number of poles with any quantum numbers, exotic in particular.

The structure and properties of the reggeon condensations may be
studied explicitly in the non-relativistic quantum mechanics with
a final-range potential. It also generates the threshold behavior
$\,\sim k^{2l}\,$ and satisfies elastic unitarity. Detailed
investigation~\cite{aas} for the Yukawa potential $V(r)=g\,
\exp(-\mu r)/r$ confirms the general character of the GP
condensations. It also shows, that poles related to bound states
(or resonances) are ``initially'' members of the set of GP
condensing poles, and ``evaporate'' from it, when attraction
increases. The energy plane for the Yukawa potential has many
Riemann sheets, and most of the energy-plane poles are ``hidden''
on remote sheets, while the poles related to bound or resonance
states approach to the physical region. The infinite number of the
Yukawa energy-plane poles can be ``visualized'' by taking the
limit $\mu\to 0$, which transforms the Yukawa potential into
Coulomb one and demonstrates the well-known accumulation of the
infinite number of Coulomb bound states near the threshold (note
that the double limiting transition $\mu\to 0,\, k\to 0$ is not
equivalent here to the similar, but reversed limit $k\to 0,\,
\mu\to 0$).

Existence of energy-plane poles with exotic quantum numbers is the
necessary condition for existence of exotic hadrons. It appears
satisfied under the familiar assumptions of unitarity and
analyticity. But only more detailed dynamics may determine whether
some of the poles emerge near the physical region, to provide
indeed the stable or resonance exotic states.

Two more ``technical'' notes may be interesting:

1) Reggeons have been used above in a non-standard manner.
Usually, to apply the CAM approach for obtaining results in the
$s$-channel (the channel where the invariant $s$ has the meaning
of the squared c.m. energy), one begins from the crossed channel,
where $t$ and $s$ are, respectively, the squared energy and
squared momentum transfer (see Ref.~\cite{col}). Analytical
continuation of partial-wave amplitudes in this $t$-channel
allows, after returning into the $s$-channel, to study behavior of
the invariant amplitude (and cross section) at the high energy $s$
and at a fixed value of the momentum transfer $t$. To obtain the
conclusion about energy-plane poles in the $s$-channel, we have
used complex angular momenta in the same $s$-channel.

2) Gribov and Pomeranchuk suggested one more situation where
infinitely many reggeons might exist near a fixed point in the
$l$-plane, this time $l=-1$~\cite{gp1}. However, accounting for
existence of the moving branch points prevents this reggeon
accumulation from emerging. The threshold GP condensations of
reggeons are not influenced by such branch points.

\section{Summary and Conclusion}

The presented results may be summarized as follows:

\begin{itemize}
\item Under the familiar assumptions of unitarity and analyticity,
hadronic amplitudes have infinite number of energy-plane poles
with any quantum numbers, both exotic and non-exotic. Thus, there
is an infinite ``reservoir'' of poles, which satisfies the
necessary condition for existence of exotics. Most of the poles
are ``hidden'' on remote Riemann sheets of the energy plane.
\item Real existence (or absence) of exotic hadrons, {\it i.e.},
of the $S$-matrix poles sufficiently near the physical region, may
be guaranteed only by more detailed dynamics.
\end{itemize}

Meanwhile, the old wisdom\cite{jj}, that ``...either these states
will be found by experimentalists or our confined, quark-gluon
theory of hadrons is as yet lacking in some fundamental, dynamical
ingredient...'', is still alive.

\section*{Acknowledgments}

I thank R.A.~Arndt, K.~Goeke, I.I.~Strakovsky, and R.L.~Workman,
my coauthors in paper~\cite{aaswg}, for stimulating discussions.
This paper and my correspondimg talks are based on the work partly 
supported by the U.S.~Department of Energy Grant DE--FG02--99ER41110, 
by the Jefferson Laboratory, by the Southeastern Universities Research 
Association under DOE Contract DE--AC05--84ER40150, by the
Russian-German Collaboration (DFG, RFBR), by the
COSY-Juelich-project, and by the Russian State grant
RSGSS-1124.2003.2. My participation in NSTAR2005 was supported by
the Organizing Committee, by the Jefferson Laboratory, and by the
George Washington University. I thank Prof. Goeke for hospitality
in the Theoretical Physics Institute II of the Ruhr-University
Bochum during writing this text.


\begin{thebibliography}{0}

\bibitem{lip} H. J. Lipkin, {\it Phys. Lett.}
{\bf 45B}, 267 (1973).
\bibitem{bur} V. Burkert, hep-ph/0510309.
\bibitem{aaswg} Ya. I. Azimov, R. A. Arndt, I. I. Strakovsky,
R. L. Workman, and K. Goeke, {\it Eur. Phys. J. } {\bf A26}, 79
(2005); hep-ph/0504022.
\bibitem{gf} V. N. Gribov, {\it Sov. Phys. JETP} {\bf 14}, 1395
(1962);\\ M. Froissart, {\it Phys. Rev.} {\bf 123}, 1053 (1961).
\bibitem{col} P. D. B. Collins, {\it An Introduction to Regge Theory
and High Energy Physics} (Cambridge Univ. Press, 1977)
\bibitem{gp} V. N. Gribov and I. Ya. Pomeranchuk, {\it Phys. Rev.
Lett.} {\bf 9}, 238 (1962).
\bibitem{az} Ya. I. Azimov, {\it Phys. Lett.} {\bf 3}, 195 (1963).
\bibitem{aas} Ya. I. Azimov, A. A. Anselm, and V. M. Shekhter,
{\it Sov. Phys. JETP} {\bf 17}, 246, 726 (1963).
\bibitem{gp1} V. N. Gribov and I. Ya. Pomeranchuk, {\it Phys. Lett.}
 {\bf 2}, 239 (1962).
\bibitem{jj} R. L. Jaffe and K. Johnson, {\it Phys. Lett.} {\bf 60B},
201 (1976).


\end{thebibliography}
\end{document}